\documentclass{iau}
\usepackage{graphicx}

\title[Spots On Pre-Main Sequence Stars] 
{The Impact of Starspots on Mass and Age \\ Estimates During The Pre-Main Sequence}

\author[Garrett Somers \& Marc H. Pinsonneault]   
{Garrett Somers \and Marc H. Pinsonneault}

\affiliation{The Ohio State University Astronomy Department\\email: {\tt somers@astronomy.ohio-state.edu}}

\pubyear{2015}
\volume{314}  
\pagerange{119--126}
\jname{Young Stars \& Planets Near the Sun}
\editors{J. H. Kastner, B. Stelzer, \& S. A. Metchev, eds.}

\def\msun   {{M$_{\odot}$}}
\def\teff   {{$T_{\rm eff}$}}
\def\teffs   {{$T_{\rm eff}$s}}
\begin{document}

\maketitle

\begin{abstract}
We investigate the impact of starspots on the evolution of late-type stars during the pre-main sequence (pre-MS). We find that heavy spot coverage increases the radii of stars by 4-10\%, consistent with inflation factors in eclipsing binary systems, and suppresses the rate of pre-MS lithium depletion, leading to a dispersion in zero-age MS Li abundance (comparable to observed spreads) if a range of spot properties exist within clusters from 3-10 Myr. This concordance with data implies that spots induce a range of radii at fixed mass during the pre-MS. These spots decrease the luminosity and \teff\ of stars, leading to a displacement on the HR diagram. This displacement causes isochrone derived masses and ages to be systematically under-estimated, and can lead to the spurious appearance of an age spread in a co-eval population.
\end{abstract}

\firstsection 
\section{Introduction}

Despite recent improvements in stellar modeling, several notable discrepancies remain between theoretical predictions of low mass stellar properties and high precision measurements from techniques such as eclipsing binary (EB) analysis and interferometry. One such discrepancy is the inflated radius problem, where young, low-mass stellar radii are observed to be larger by $5-10$\% than theoretical predictions. A second discrepancy is between the observed lithium patterns of young clusters such as the Pleiades, which host abundance dispersions in excess of an order-of-magnitude at fixed \teff, and standard model theoretical predictions, which anticipate no dispersion at fixed \teff. Notably, the most Li rich stars in the Pleiades are also the most rapidly rotating, the opposite sense of the correlation expected from rotational mixing.

In two recent papers, we suggested that if some mechanism induced inflated radii in rapidly rotating, low-mass stars during the pre-MS, then the Li destruction rate in the inflated stars would be suppressed, and one could explain both the inflated radius problem and the Li-rotation correlation in the Pleiades, with a single mechanism (\cite[Somers \& Pinsonneault 2014]{somers14}; \cite[Somers \& Pinsonneault 2015a]{somers15a}). In this proceedings, we summarize results from our recent work which explores this possibility using a specific inflation mechanism, starspots.

\section{Starspots}

Starspots are the visual manifestation of concentrated magnetic fields near the surfaces of stars. Spots inhibit energy transport through convection, and alter the pressure conditions at the photosphere, inducing a global structural response that can be modeled (\cite[Gough \& Taylor 1966]{gough66}). Spots have long been argued to impact the luminosities, temperatures, and radii of stars (e.g. \cite[Spruit \& Weiss 1986]{spruit86}), but in only a few cases has their effect on the pre-main sequence (pre-MS) been examined (e.g. \cite[Jackson \& Jeffries 2014]{jackson14}), and never with a modern stellar evolution code.

Young, active stars are typically the most heavily spotted, so a systematic study of the structural impact of spots during the pre-MS and early MS may prove crucial to our understanding of early stellar evolution. To this end, we have incorporated in our stellar evolution code a treatment of spots which accounts for two relevant physical processes: 1) the redistribution of flux in the interior, modeled by explicitly altering the radiative gradient; 2) the altered boundary conditions at the surface, modeled by self-consistently solving for the pressure in the un-spotted regions (the full details are presented in $\S 2.2$ of \cite[Somers \& Pinsonneault 2015b]{somers15b}).

Using this method, we calculated a grid of low mass (0.1-0.8\msun) and solar mass (0.8-1.2\msun) stars, with various spot properties. Our spotted models each have a spot$-$photosphere temperature contrast of 0.8, and host a variety of spot filling factors up to 50\% surface coverage, the upper limit of claimed properties in the literature. 

\section{Impact on Stellar Radii and Lithium Abundances}

\begin{figure}[b]
\begin{center}
\includegraphics[width=2.6in]{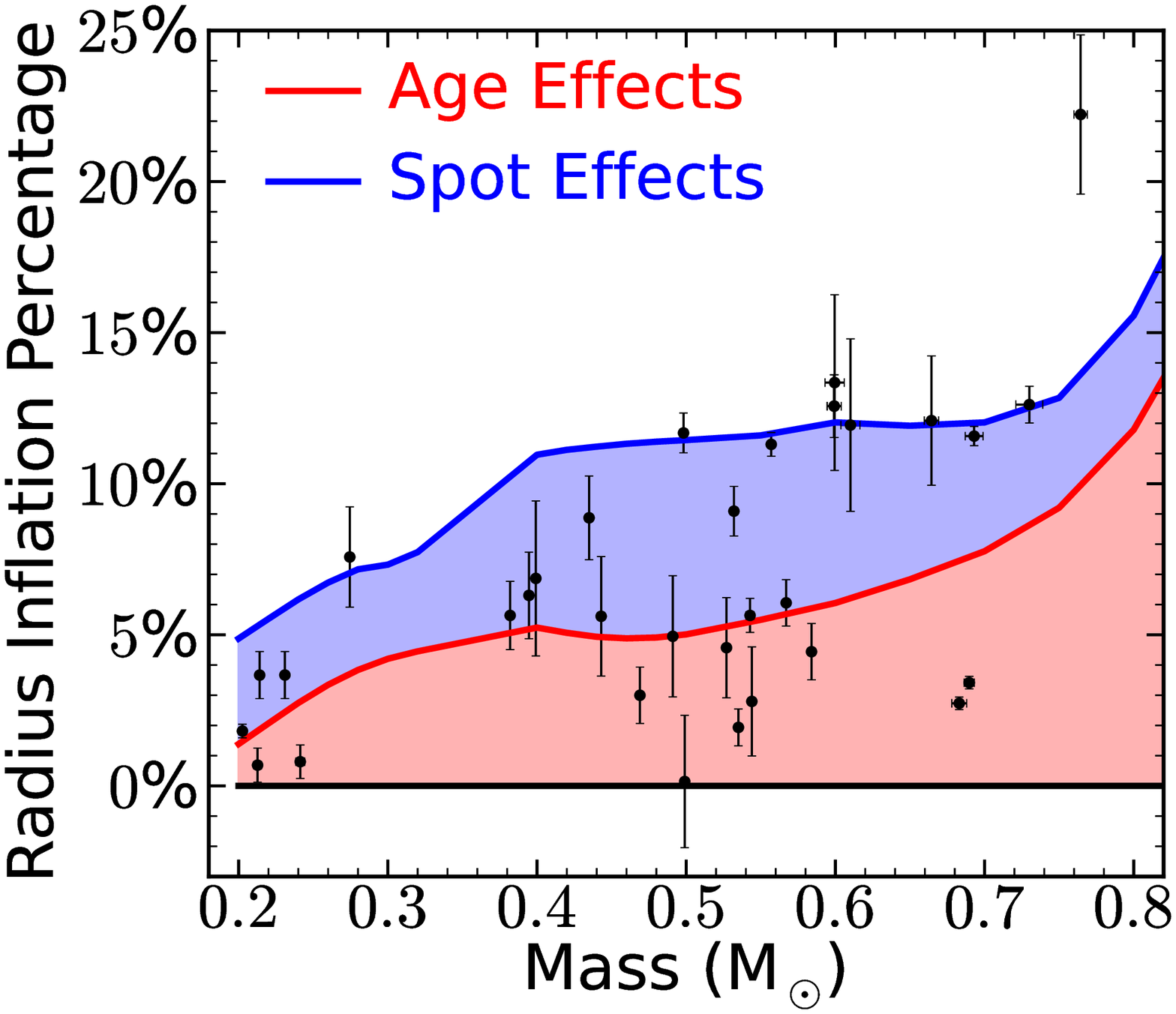} 
\includegraphics[width=2.5in]{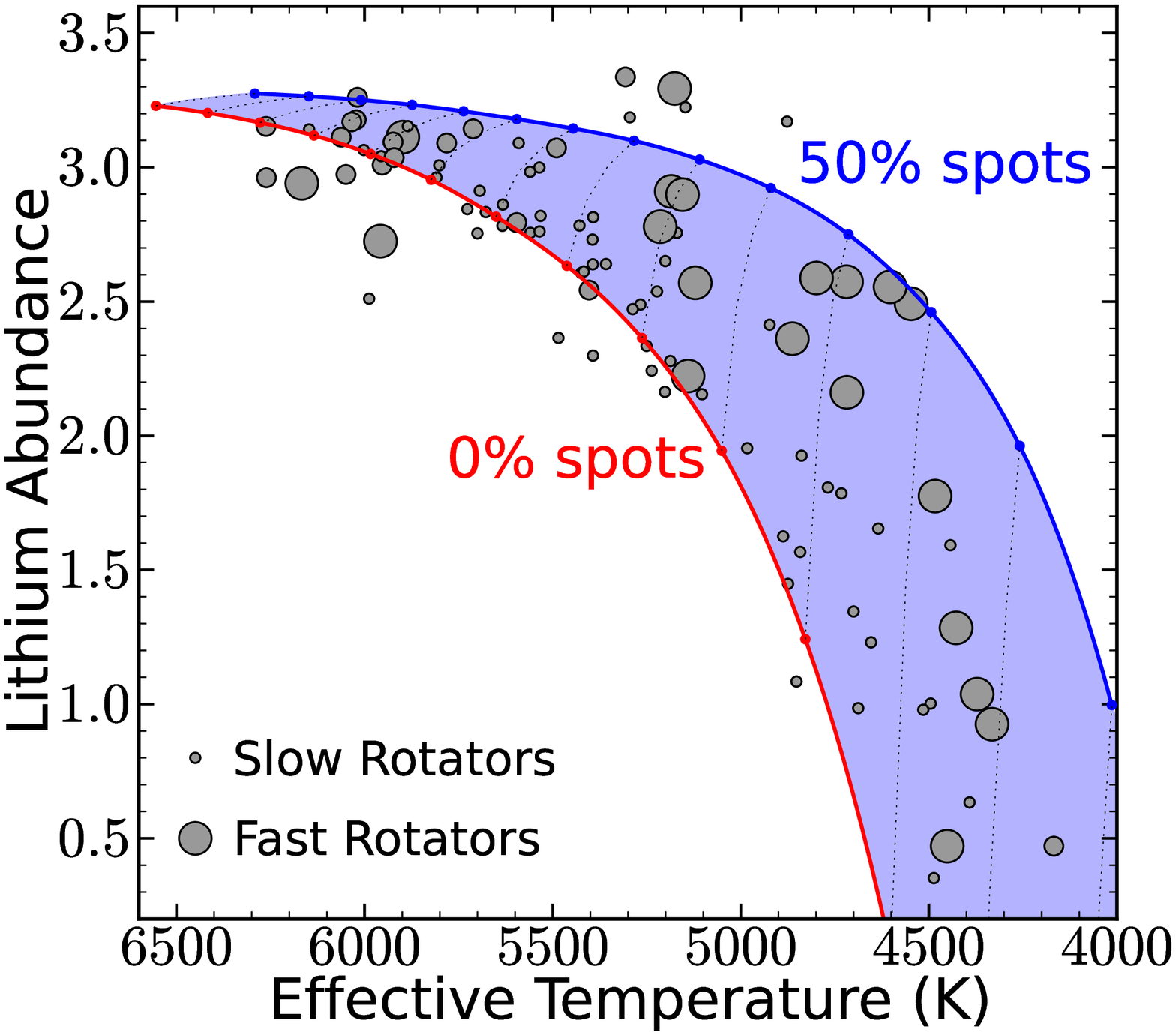} 
 \caption{\textit{Left:} The impact of spot and age effects on stellar radii. Each point shows the radius inflation inferred for an eclipsing binary member relative to our models (data from \cite[Torres et al. 2010]{torres10} and \cite[Feiden \& Chaboyer 2012]{feiden12}). The red region shows the radius increase incurred over 10 Gyr, and the blue region shows the combined impact of heavy spot coverage and age. With both effects considered, nearly all the data can be explained. \textit{Right:} The impact of spots on Li destruction during the pre-MS. Spot coverage up to 50\% can suppress Li destruction enough to explain the Li spread at fixed \teff\ observed in the Pleiades (grey circles; Li from \cite[Soderblom et al. 1993]{soderblom93}, rotation from \cite[Hartman et al. 2010]{hartman10}).}
   \label{fig1}
\end{center}
\end{figure}

We first explored the impact of spots on stellar radii. We find that heavy spot coverage leads to a substantial enhancement at all ages, developing on the early pre-MS and persisting onto the MS. For all masses in our range, radius anomalies peak at 8-12\% during the pre-MS, and converge to a mass dependent pattern at the ZAMS, where fully convective stars are inflated by $\sim$4\%, and solar mass stars are inflated by up to 10\%. The larger anomalies on the pre-MS result primarily from the reduced Hayashi contraction rate of spotted stars, a consequence of their suppressed luminosity ($\S$\ref{sec4}). By contrast, MS anomalies result predominately from the impact of spots on the pressure boundary conditions at the surface, and at the transport of flux in the upper envelope.

In the left panel of Fig. \ref{fig1}, we compare the radius predictions of our spotted models to EB masses and radii. Each data point shows the fractional inflation percentage relative to our zero-age MS predictions. The red region shows the impact of 10 Gyrs of stellar evolution on the radius, indicating that a subset of the data could be explained solely by age effects, though more than half cannot. The blue region shows the area permitted when including both spot and age effects, which could both be at work in short-period, (presumably) tidally-locked EBs. With only one exception, our spot models can account for the anomalous radii of the comparison sample. This demonstrates that our models predict reasonable inflation factors for observationally motivated spot properties.

Next, we explore the destruction of lithium in our models. Li burns at $T \sim 2.5 \times 10^6$~K, and is depleted from stellar surfaces during the pre-MS until a substantial radiative core has formed. Standard models (red line in the right panel of Fig. \ref{fig1}) predict that the mass and composition of a star uniquely determine its Li destruction during the MS, and consequently predict no spread in Li at fixed \teff\ within clusters. By contrast, a large dispersion in Li abundance at fixed \teff\ (grey points) is observed in the 120 Myr Pleiades cluster, suggesting additional mechanisms are at play.

Spotted models are physically larger during the pre-MS, leading to a reduction in the temperature at the base of the surface convection zone, and thus a suppressed rate of Li destruction. The blue line in the right panel of Fig. \ref{fig1} shows the Li abundances resulting from a constant filling factor of 50\% during the pre-MS, and the blue shaded regions show abundances produced by intermediate spot properties. The majority of the data fall within this range, suggesting that stars of equal mass must have dissimilar radii during the pre-MS to explain the data. Furthermore, the most Li-rich stars are the fastest spinning, suggesting that this radius dispersion arises from a physical process linked to rotation $-$ spots are a prime candidate. 

Collectively, Li and radius data suggest that a range of radii at fixed mass, driven by magnetic activity, emerge in open clusters during the pre-MS and persist onto the MS. We now discuss an important implication of this conclusion.

\section{Consequences for Inferred Masses and Ages}\label{sec4}

\begin{figure}[t]
\begin{center}
\includegraphics[width=2.5in]{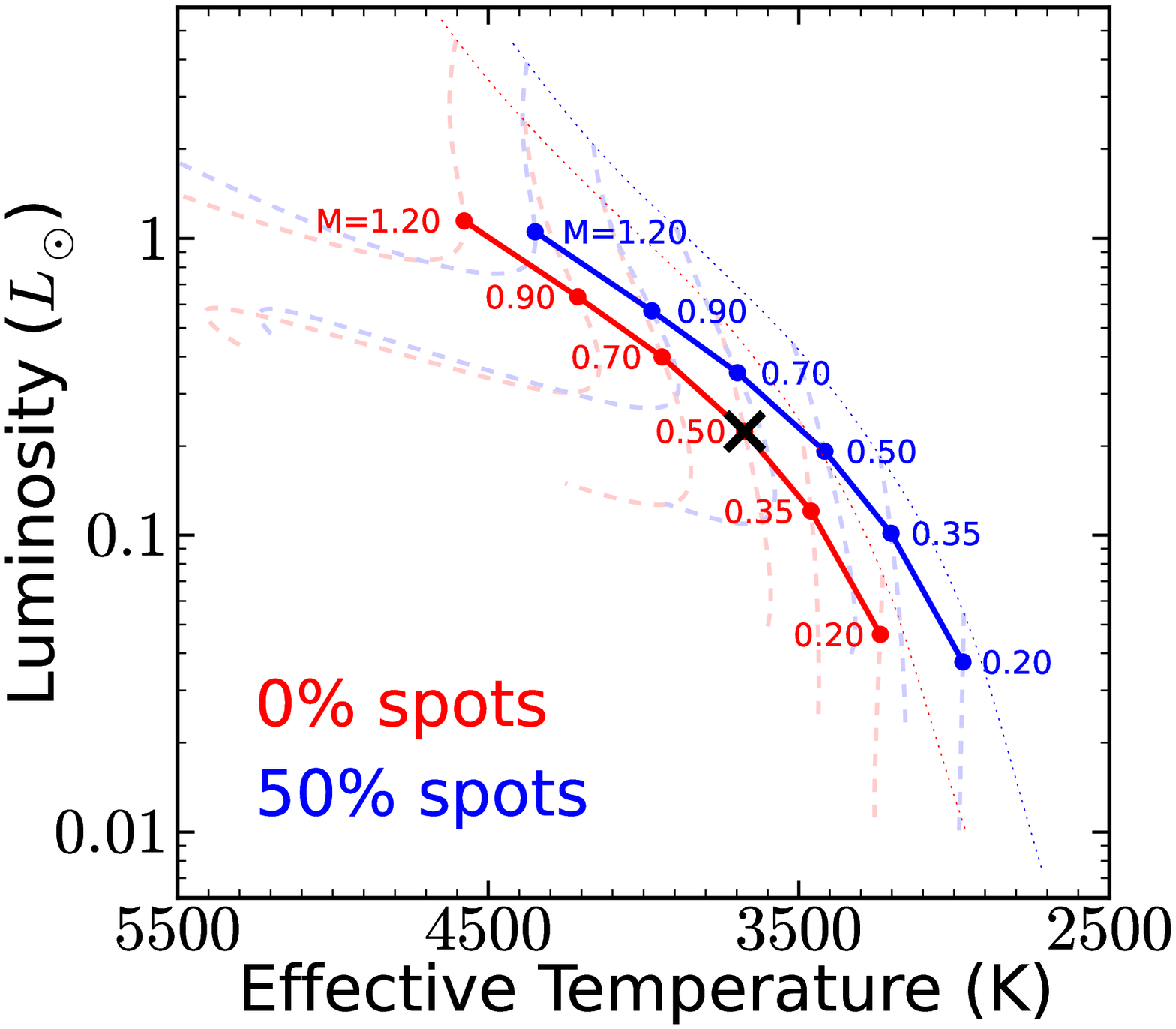} 
\includegraphics[width=2.5in]{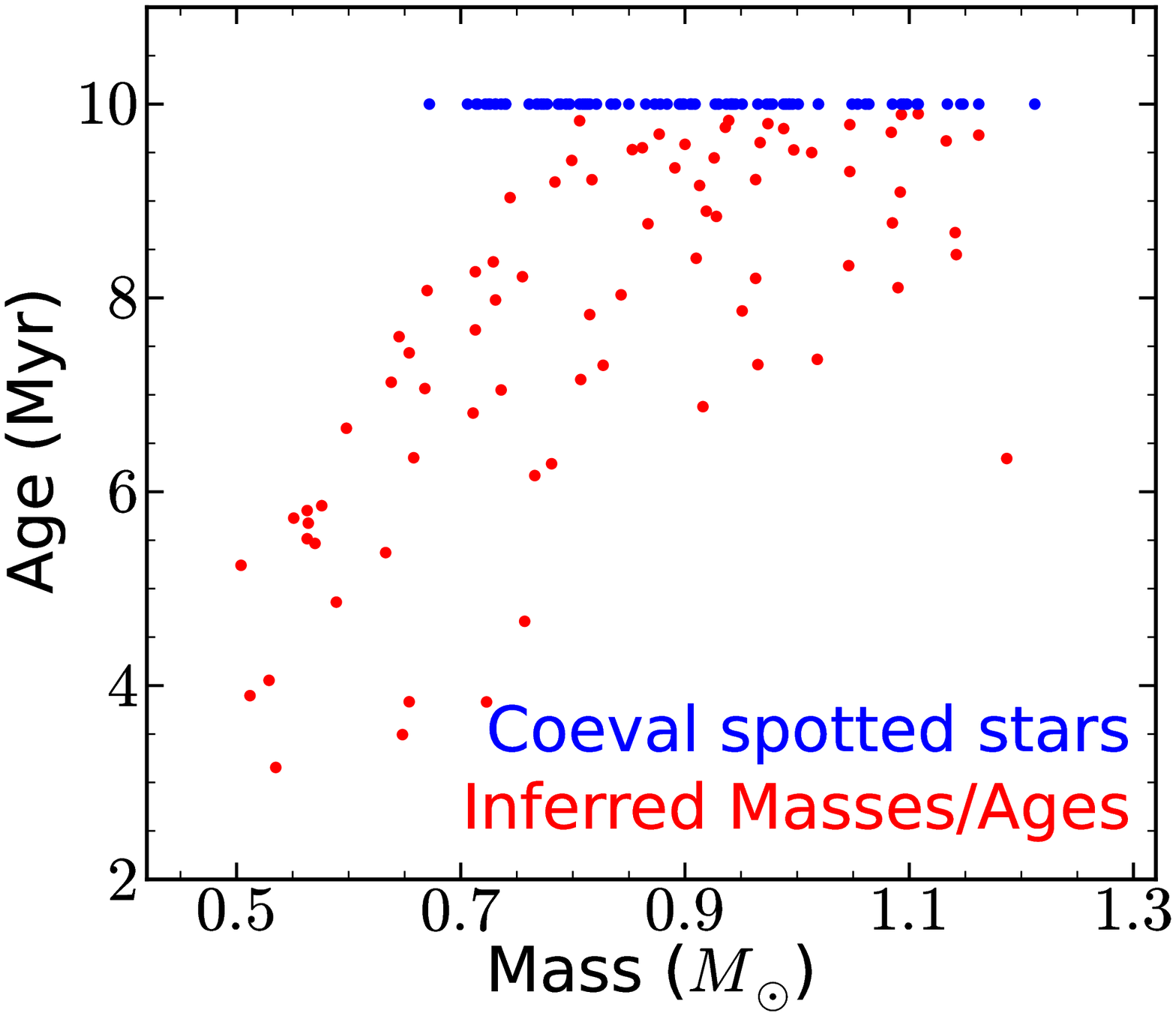} 
 \caption{\textit{Left:} 3 Myr isochrones for un-spotted (red) and spotted (blue) models. A pre-MS star at the location of the black cross will be incorrectly interpreted if its spot properties are not accounted for. \textit{Right:} Mass and age errors incurred when inferring properties of a spotted population with standard isochrones.}
   \label{fig2}
\end{center}
\end{figure}

The left panel of Fig. \ref{fig2} shows standard (red) and 50\% spotted (blue) isochrones at 3 Myr in the HR diagram. The blue isochrone is displaced downward and to the right, indicating that spots reduce the luminosity and \teff\ of their host stars. Now consider the location of the black cross. If a pre-MS stars is observed at this HR-diagram location, one might infer a mass of 0.5\msun\ and an age of 3 Myr through comparison with the standard isochrones. However, if this star is actually spotted, then the blue isochrones should be used to derive its properties, in which case one finds $M \sim 0.75$\msun, and $t \sim 8$~Myr. This demonstrates that spotted stars are \textit{systematically older and more massive than often measured.}

To estimate the magnitude of this effect, we return to the Pleiades Li abundances in Fig. \ref{fig1}. If the cause of the Li dispersion is indeed pre-MS spots, then the current Li abundance of each individual star tells us what its spot properties must have been on the pre-MS. We can therefore derive the distribution of pre-MS spot properties that the Pleiades \textit{must have had}, in order to produce the current \teff\ vs. Li distribution. We can then ask what masses and ages \textit{would we have measured} from this distribution, using standard isochrones. The results are in the right panel of Fig. \ref{fig2}. The blue points show the masses derived from the Pleiades, backwards modeled to a coeval age of 10 Myr. The red points show the masses and ages that isochrone fitting gives for this population. As can be seen, there is a small systematic trend towards lower inferred masses, and a very strong trend towards lower ages, with errors reaching a factor of $3 \times$ for the most vulnerable stars. This is not a subtle effect! We further note that the lowest mass stars appear to be the youngest, a feature which has been observed in some young clusters (e.g. \cite[Herczeg \& Hillenbrand 2015]{herczeg15}). This suggests that many young associations may in fact be older than often quoted, that age spreads claimed in some clusters may be spurious, and that age errors are largest for the lowest mass stars.

\section{Conclusions}

We have found that the impact of spots on stellar models is to increase their radii, and to lower their luminosities and \teffs, in a mass dependent fashion. The increased radii due to spots can explain the observed radius inflation of some low mass stars, and can account for the suppressed Li depletion in rapid rotators during the pre-MS. These data collectively suggest that stars of equal mass have different physical sizes at fixed age around 3-10 Myr. As a consequence, masses and ages derived with isochrones will be erroneously low and young, respectively, and may create the appearance of age spreads in co-eval populations.


\begin{thebibliography}{}

\bibitem[Feiden \& Chaboyer(2012a)]{feiden12}
{Feiden, G.~A., \& Chaboyer, B.} 2012,
\textit{ApJ}, 757, 42 

\bibitem[Gough \& Tayler(1966)]{gough66}
{Gough, D.~O., \& Tayler, R.~J.} 1966,
\textit{MNRAS}, 133, 85 

\bibitem[Hartman et al.(2010)]{hartman10}
{Hartman, J.~D., Bakos, G.~{\'A}., Kov{\'a}cs, G., \& Noyes, R.~W.} 2010,
\textit{MNRAS}, 408, 475 

\bibitem[Herczeg \& Hillenbrand(2015)]{2015arXiv150506518H}
{Herczeg, G.~J., \& Hillenbrand, L.~A.} 2015,
arXiv:1505.06518 

\bibitem[Jackson \& Jeffries(2014a)]{jackson14}
{Jackson, R.~J., \& Jeffries, R.~D.} 2014,
\textit{MNRAS}, 441, 2111 

\bibitem[Soderblom et al.(1993)]{soderblom93}
{Soderblom, D.~R., Jones, B.~F., Balachandran, S., et al.} 1993,
\textit{AJ}, 106, 1059 

\bibitem[Somers \& Pinsonneault(2014)]{somers14}
{Somers, G., \& Pinsonneault, M.~H.} 2014,
\textit{ApJ}, 790, 72 

\bibitem[Somers \& Pinsonneault(2015a)]{somers15a}
{Somers, G., \& Pinsonneault, M.~H.} 2015a,
\textit{MNRAS}, 449, 4131 

\bibitem[Somers \& Pinsonneault(2015b)]{somers15b}
{Somers, G., \& Pinsonneault, M.~H.} 2015b,
arXiv:1506.01393 

\bibitem[Spruit \& Weiss(1986)]{spruit86}
{Spruit, H.~C., \& Weiss, A.} 1986,
\textit{AAP}, 166, 167 

\bibitem[Torres et al.(2010)]{torres10}
{Torres, G., Andersen, J., \& Gim{\'e}nez, A.} 2010,
\textit{AAPR}, 18, 67 

\end{thebibliography}
\end{document}